\documentclass[11pt,a4paper]{article}

\usepackage{amsmath}
\usepackage{amsthm,amssymb}

\usepackage{graphicx}
\usepackage{grffile}
\usepackage{subfigure}
\usepackage[normalem]{ulem}

\usepackage{tikz}
\usetikzlibrary{scopes}

\newcommand{\beq}{\begin{equation}}
\newcommand{\eeq}{\end{equation}}

\newcommand{\<}{\langle}
\renewcommand{\>}{\rangle}

\DeclareMathOperator*{\argmax}{arg\,max}

\numberwithin{equation}{section}

\begin{document}
\title{Semi-flexible interacting self-avoiding trails on the square lattice}
\author{A Bedini$^1$, A L Owczarek$^1$ and T Prellberg$^2$\\
  \footnotesize
  \begin{minipage}{13cm}
    $^1$ Department of Mathematics and Statistics,\\
    The University of Melbourne, Parkville, Vic 3010, Australia.\\
    \texttt{\{abedini,owczarek\}@unimelb.edu.au}\\[1ex] 
$^2$ School of Mathematical Sciences\\
Queen Mary University of London\\
Mile End Road, London E1 4NS, UK\\
\texttt{t.prellberg@qmul.ac.uk}
\end{minipage}
}

\maketitle  

\begin{abstract}
  Self-avoiding walks self-interacting via nearest neighbours (ISAW)
  and self-avoiding trails interacting via multiply-visited sites
  (ISAT) are two models of the polymer collapse transition of a
  polymer in dilute solution. On the square lattice it has been
  established numerically that the collapse transition of each model
  lies in a different universality class.

  It has been shown that by adding stiffness to the ISAW model a
  second low temperature phase eventuates and a more complicated phase
  diagram ensues with three types of transition that meet at a
  multi-critical point. For large enough stiffness the collapse
  transition becomes first-order. Interestingly, a phase diagram of a
  similar structure has been seen to occur in an extended ISAT model
  on the triangular lattice without stiffness.  It is therefore of
  interest to see the effect of adding stiffness to the ISAT model.

  We have studied by computer simulation a generalised model of
  self-interacting self-avoiding trails on the square lattice with a stiffness parameter
  added. Intriguingly, we find that stiffness does not change the order
  of the collapse transition for ISAT on the square lattice for a very wide range of stiffness weights. 
  While at the lengths considered there are clear bimodal distributions for very large stiffness,
  our numerical evidence strongly suggests that these are simply finite-size effects associated with
  a crossover to a first-order phase transition at infinite stiffness.
\end{abstract}

\newpage

\section{Introduction}
\label{sec:introduction}

The collapse transition of a polymer in a dilute solution has been a
continuing focus of study in lattice statistical mechanics for decades
\cite{gennes1975a-a,gennes1979a-a}. This transition describes the
change in the scaling of the polymer with length that occurs as the
temperature is lowered. At high temperatures the radius of gyration of
polymer scales in a way swollen relative to a random walk: this is
known as the excluded volume effect. At low temperatures a polymer
condenses into dense, usually disordered, globule, with a much smaller
radius of gyration.  The interest in this phase transition has
occurred both because of the motivation of physical systems but also
because of the study of integrable cases
\cite{nienhuis1982a-a,warnaar1992b-a} of lattice models, that have proved
especially fruitful in two dimensions.  While the canonical lattice
model of the configurations of a polymer in solution has been the
self-avoiding walk (SAW), where a random walk on a lattice is not
allowed to visit a lattice site more than once, an alternative has
been to use bond-avoiding walks, or a self-avoiding trail. A
self-avoiding trail (SAT) is a lattice walk configuration where the
excluded volume is obtained by preventing the walk from visiting the
same bond, rather than the same site, more than once. These were used
initially to model polymers with loops \cite{malakis1976a-a} but have
subsequently occurred in integrable loop models in two dimensions
\cite{warnaar1992b-a}.  A model of collapsing polymers can be
constructed starting from self-avoiding trails, known as interacting
self-avoiding trails (ISAT). Here energies are associated with
multiply-visited sites and by favouring configurations with many such
sites a collapse transition can be initiated.

Owczarek and Prellberg studied numerically the ISAT collapse on the
square lattice by two different approaches
\cite{owczarek1995a-:a,owczarek2007c-:a} and in either case found a
strong continuous transition with specific heat exponent $\alpha =
0.81(3)$. Recently, on the triangular lattice Doukas \textit{et al.\
}\cite{doukas2010a-:a} found that by changing the weighting of doubly
and triply visited sites a first-order transition can ensure or
alternatively, depending on the ratio of these weightings, a weaker
second-order transition that mimics the collapse found in the
canonical interacting self-avoiding walk (ISAW) model (also know as
the $\theta$-point). They also found that the low temperature phase
could become fully dense rather than globular.

Coorespondingly, there is also a modification of the ISAW model that displays two
phase transitions for a range of parameters, namely the
\emph{semi-flexible} ISAW model
\cite{bastolla1997a-a,vogel2007a-a,doye1997a-a}. Here two energies
are included: the nearest-neighbour site interaction of the ISAW model
and also a stiffness energy associated with consecutive parallel bonds
of the walk (equivalently, a bending energy for bends in the walk). This
has been studied on the cubic lattice by Bastolla and Grassberger
\cite{bastolla1997a-a}. 
They showed that when there is a strong energetic preference for straight segments,
this model undergoes a single first-order transition from the
excluded-volume high-temperature state to a fully dense state.
On the other hand, if there is only a weak preference for straight
segments, the polymer undergoes two phase transitions. On lowering the
temperature the polymer undergoes a $\theta$-point transition to
the liquid globule followed by a first-order
transition to the fully dense phase at a lower temperature. Recent work by Krawczyk \textit{et al.} \cite{krawczyk2010a-:a} concerning  the ISAW model on the square lattice in
presence of a stiffness parameter  showed that the introduction of
stiffness can change the universality class of the collapse
transition in two dimensions. For large stiffness the transition becomes first order and
the collapsed phase moves from being globular to fully dense.

Recently, Foster \cite{foster2011a-a} introduced and studied a generalised ISAT model
on the square lattice which incorporates stiffness.
Using transfer matrices and the phenomenological renormalisation group, that study 
predicted that the ISAT universality class is
unaffected by a range of values of stiffness. However, the results
suggested the appearance of a first-order transition for sufficiently
large stiffness.

In this work we use Monte Carlo simulation to explore ISAT in presence
of stiffness and the predictions of Foster \cite{foster2011a-a}. We
also explore the low temperature phase of the model and find that there is only one low temperature phase and that it is fully dense for the range of stiffness studied. 
 
\section{ISAT}
\label{sec:isat}

The model of interacting trails on the square lattice is defined as
follows. Consider the ensemble $\mathcal T_n$ of self-avoiding trails
(SAT) of length $n$, that is, of all lattice paths of $n$ steps that
can be formed on the square lattice such that they never visit the
same bond more than once.
Given a SAT $\psi_n \in \mathcal T_n$, we associate an energy
$-\varepsilon_t$ with each doubly visited site, and denote their number by $m(\psi_n)$.
The probability of $\psi_n$ is then given by
\begin{equation}
	\frac{e^{\beta \varepsilon_t m(\psi_n)}}{Z^{ISAT}_n(T)},
\end{equation}
where we define the Boltzmann weight $\omega_t = \exp(\beta
\varepsilon_t)$ and $\beta$ is the inverse temperature $1/k_BT$. The
partition function of the ISAT model is given by
\begin{equation}
	Z^{ISAT}_n(T) = \sum_{\psi_n\in \mathcal T_n}\ \omega_t^{m(\psi_n)}.
\end{equation}
The finite-length reduced free energy is
\begin{equation}
	\kappa_n(T) = \frac{1}{n} \log\ Z_n(T)
\end{equation}
and the thermodynamic limit is obtained by taking the limit of large $n$, i.e.,
\begin{equation}
	\kappa(T) = \lim_{n \to \infty} \kappa_n(T).
\end{equation}
It is expected that there is a collapse phase transition at a
temperature $T_c$ characterized by a non-analyticity in $\kappa(T)$.

The collapse transition can be characterized via a change in the
scaling of the size of the polymer with temperature. It is expected
that some measure of the size, such as the radius of gyration or the
mean squared distance of a monomer from the end points, $R_n^2(T)$,
scales at fixed temperature as
\begin{equation}
  R_n^2(T) \sim A n^{2\nu}
\end{equation}
with some exponent $\nu$. At high temperatures the polymer is swollen
and in two dimensions it is accepted that $\nu=3/4$
\cite{nienhuis1982a-a}. At low temperatures the polymer becomes dense
in space, though not necessarily space filling, and the exponent is $\nu=1/2$. 
However, for the ISAT model the collapsed phase has been seen to be space filling
\cite{bedini2012=a-:a}.
If the collapse transition is second-order, the scaling at
$T_c$ of the size is intermediate between the high and low temperature
forms. In the thermodynamic limit the expected 
singularity in the free energy can be seen in its second
derivative (the specific heat). Denoting the (intensive) finite length
specific heat \emph{per monomer} by $c_n(T)$, the thermodynamic limit
is given by the long length limit as
\begin{equation}
  C(T) = \lim_{n\rightarrow\infty} c_n(T)\;.
\end{equation}
One expects that the singular part of the specific heat behaves as
\begin{equation}
  C(T) \sim B |T_c -T|^{-\alpha}\;,
\end{equation}
where $\alpha<1$ for a second-order phase transition.  The singular
part of the thermodynamic limit internal energy behaves as
\begin{equation}
  U(T) \sim B |T_c -T|^{1-\alpha}\;,
\end{equation}
if the transition is second-order, and there is a jump in the internal
energy if the transition is first-order (an effective value of
$\alpha=1$).

Moreover, one expects crossover scaling forms \cite{brak1993a-:a} to
apply around this temperature, so that
\begin{equation}
  c_n(T) \sim n^{\alpha\phi} \; {\cal C}((T - T_c)n^\phi)
\label{spec-heat-scaling}
\end{equation}
with  $0<\phi < 1$ if the transition is second-order, and 
\begin{equation}
  c_n(T) \sim n \; {\cal C}((T - T_c)n)
\end{equation}
if the transition is first-order. From \cite{brak1993a-:a} we point
out that the exponents $\alpha$ and $\phi$ are related via
\begin{equation}
  2-\alpha = \frac{1}{\phi}\;.
\end{equation}
Important for numerical estimation is the use of
equation~(\ref{spec-heat-scaling}) at the peak value of the specific
heat given by $y^{peak}= (T - T_c)n^\phi$ so that
\begin{equation}
  c^{peak}_n(T) \sim   {\cal C}^{peak} \; n^{\alpha\phi}
\label{spec-heat-peak-scaling}
\end{equation}
where ${\cal C}^{peak}= {\cal C}(y^{peak})$ is a constant.

A previous study \cite{owczarek2007c-:a} of ISAT model on the square
lattice has shown that there is a collapse transition with a strongly
divergent specific heat, with
\begin{equation}
  \alpha\phi =0.68(5)
\label{isat-exponent}
\end{equation}
and so the individual exponents have been estimated as
\begin{equation}
  \phi =0.84(3)\quad \mbox{ and } \quad \alpha=0.81(3)\;. 
\end{equation}
At $T=T_c$, given by \cite{meirovitch1989d-a,bradley1990a-a}
\begin{equation}
\omega_t=3,
\end{equation}
it was  predicted \cite{owczarek1995a-:a} that
\begin{equation}
  R_n^2(T) \sim A n\left(\log n\right)^2\;.
\end{equation}

\section{Semi-flexible ISAT}
\label{sec:semi-flex-isat}

\begin{figure}[ht!]
  \centering
  \begin{tikzpicture}
    \draw[help lines] (-1,0) grid (6,3); \fill (0,0) circle (2pt);
    \draw[->,very thick,rounded corners=5pt] (0,0) -- ++(0,2) --
    ++(1,0) node [below left=2pt] {$\tau$} -- ++(0,-1) -- ++(1,0) --
    ++(0,-1) -- ++(1,0) -- ++(0,1) node [below right=1pt] {$\tau p^2$} --
    ++(0,2) -- ++(2,0) -- ++(0,-1)
    node[circle=2pt,draw,thin,pin={[pin edge={black,<-}]50:$p$}] {} --
    ++(0,-1) -- ++(-3,0) -- ++(0,1) -- ++(-1,0) -- ++(0,1) ;
  \end{tikzpicture}
  \caption{An example of semi-flexible ISAT configuration with three
    ($m = 3$) multiple-visit interactions associated with the
    Boltzmann weight $\tau$ and seven ($s = 7$) straight segments each
    associated with the Boltzmann weight $p$. Note that when the trail
    crosses itself, the crossing site is associated with a total weight
    $\tau p^2$ as a crossing necessarily requires two straight segments.}
  \label{fig:semi-flex-isat}
\end{figure}

The semi-flexible ISAT (SFISAT) model can be defined as follows.
Consider the set of bond-avoiding paths $\mathcal T_n$ as defined in
the previous section. Given a SAT $\psi_n \in \mathcal T_n$, we
associate an energy $-\varepsilon_t$ every time the path visits the
same site more than once, as in ISAT. Additionally, we define a straight
segment of the trail by two consecutive parallel edges, and we associate an
energy $-\varepsilon_s$ to each straight segment of trail.

For each configuration $\psi_n \in \mathcal T_n$ we count the number
$m(\psi_n)$ of doubly-visited sites and $s(\psi_n)$ of straight
segments: see Figure~\ref{fig:semi-flex-isat}. Hence we associate with
each configuration a Boltzmann weight $\tau^{m(\psi_n)} p^{s(\psi_n)}$
where $\tau = \exp(\beta \varepsilon_t)$, $p = \exp(\beta
\varepsilon_s)$, and $\beta$ is the inverse temperature $1/k_B T$. The
partition function of the SFISAT model is given by
\begin{equation}
  Z_n(\tau, p) = \sum_{\psi_n\in\mathcal T_n}\
  \tau^{m(\psi_n)} p^{s(\psi_n)}
  . 
\end{equation}
The probability of a configuration $\psi_n$ is then
\begin{equation}
  p(\psi_n; \tau, p) = \frac{ \tau^{m(\psi_n)}
    p^{s(\psi_n)} }{ Z_n(\tau, p) }
  .
\end{equation}
When we set $p = 1$ the trail is fully flexible and the model reduces
to the ISAT model. On the other end, if we set $p = 0$ straight
segments are excluded and our model requires the path to turn at every site: this is known as the ``L-lattice''. Trails on the L-lattice may be mapped \cite{bradley1989a-a} into the Interacting
Self-Avoiding Walk model on the Manhattan lattice \cite{prellberg1994a-:a}. As a consequence of the mapping, the transition of ISAT on the L-lattice has been shown to be $\theta$-like with a convergent specific heat, unlike ISAT on the square lattice.

The average of any quantity $Q$ over the ensemble set of path
$\mathcal T_n$ is given generically by
\begin{equation}
  \langle Q \rangle(n; \tau, p) = \sum_{\psi_n\in\mathcal
    T_n} Q(\psi_n) \, p(\psi_n; \tau, p)
  .
\end{equation}
In particular, we can define the average number of doubly-visited
sites per site and their respective fluctuations as
\begin{equation}
  u = \frac{ \langle m \rangle }{n}
  \quad \mbox{ and } \quad
  c = \frac{ \langle m^2 \rangle - \langle m
    \rangle^2 }{n}
  .
\end{equation}
One can also consider the average number of straight sections of the trail and their fluctuations
\begin{equation}
  u^{(s)} = \frac{ \langle s \rangle }{n}
  \quad \mbox{ and } \quad
  c^{(s)} = \frac{ \langle s^2 \rangle - \langle s
    \rangle^2 }{n}
  .
\end{equation}
An important quantity for what follows is the proportion of the sites
on the trail that are at lattice sites which are not doubly occupied:
\begin{equation}
  v_n = 1 - \frac{2 \< m \>}{n}
  .
\end{equation}
Foster \cite{foster2011a-a} predicted that the universality class of
fully flexible ISAT at $p=1$ extends to other values of $p$ and
also that for large $p$ there may be a change to a first-order
transition.

Defining the collapse as occurring at $\tau_c(p)$ for constant $p$, it follows that
\begin{equation}
\tau_c(1)=3 \qquad \text{ and } \qquad \tau_c(0) =2
\end{equation}
from the ISAT and L-lattice \cite{bradley1989a-a,meirovitch1989d-a,bradley1990a-a} results.

\section{Results}
\label{sec:results}

We began by simulating the full two parameter space by using the
flatPERM algorithm \cite{prellberg2004a-a}. FlatPERM outputs an
estimate $W_{n,\mathbf{k}}$ of the total weight of the walks of length
$n$ at fixed values of some vector of quantities
$\mathbf{k}=(k_1,k_2,\dotsc,k_{\ell})$. From the total weight one can
access physical quantities over a broad range of temperatures through
a simple weighted average, e.g.
\begin{align}
  \< \mathcal O \>_n(\tau) = \frac{\sum_{\mathbf{k}} \mathcal O_{n,\mathbf{k}}\,
   \left(\prod_j \tau_j^{k_j}\right) \, W_{n,\mathbf{k}}}{\sum_\mathbf{k} \left(\prod_j \tau_j^{k_j}\right) \, W_{n,\mathbf{k}}}.
\end{align}
The quantities $k_j$ may be any subset of the physical parameters of
the model. In our case we begin by using $k_1=m$ and $k_2=s$.

We have simulated SFISAT using the full two-parameter flatPERM
algorithm up to length $n = 512$, with $10^5$ iterations, collecting
$7.8 \cdot 10^9$ samples at the maximum length.

\begin{figure}[ht!] \centering
  {\includegraphics{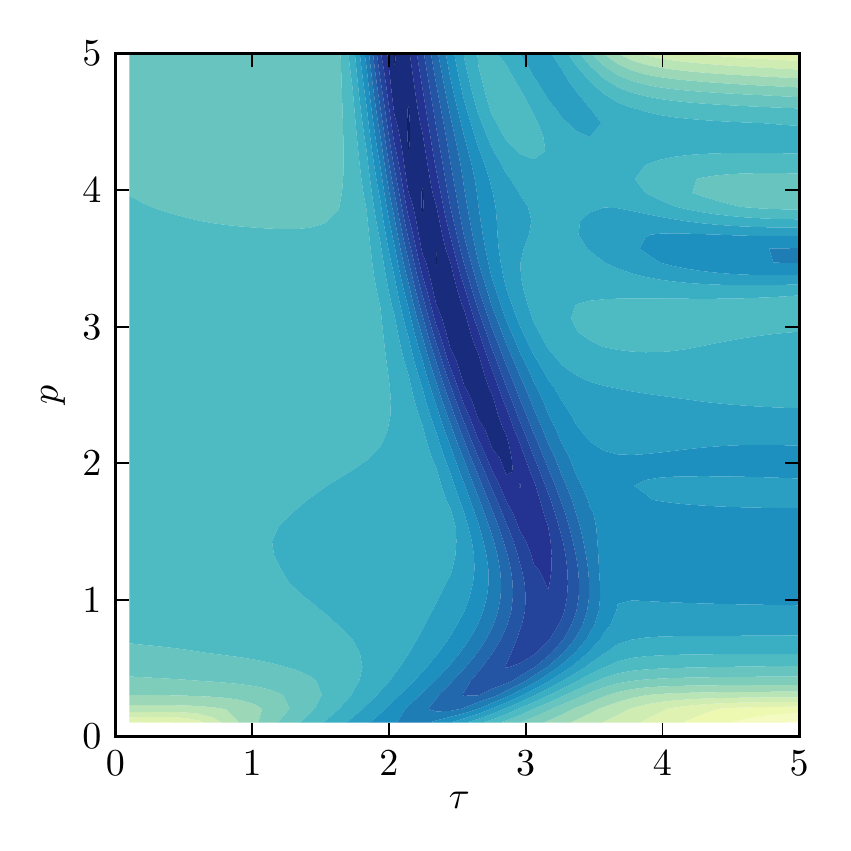}}
  \caption{Density plot of the logarithm of the largest eigenvalue
    $\lambda_{max}$ of the matrix of second derivatives of the free
    energy with respect to $\tau$ and $p$ at length $512$. Darker regions
    denote larger values.}
  \label{fig:eigenvalue_plot}
\end{figure}

To obtain a landscape of possible phase transitions we plot the largest
eigenvalue of the matrix of second derivatives of the free energy  with
respect $\tau$ and $p$ (measuring the fluctuations and covariance in $m$ and $s$)  at length $n=512$ in
Figure~\ref{fig:eigenvalue_plot}.

We notice that the strong peak corresponding to the ISAT transition at
$(\tau,p)=(3,1)$ extends upward to larger values of $p$,
becoming stronger as $p$ increases, as predicted in \cite{foster2011a-a}. At smaller values of $p$ the
peak of the specific heat seems to get weaker as it reaches the point
$(\tau,)=(2,0)$. At this point, corresponding to a $\theta$-like
transition, the specific heat is known to not diverge.

To investigate the nature of the transition at larger values of the
stiffness parameter $p$, we have run extensive simulations of the model at
few fixed values of $p$ all up to length $n = 1024$. We have simulated
different values of $p$ between $p = 0.1$ and $p = 10$, running
between $S \simeq 2.3 \cdot 10^6$ and $S \simeq 4 \cdot 10^6$
iterations, and collecting between $6.0 \cdot 10^9$ and $1.7 \cdot
10^{10}$ samples at the maximum length. Following
\cite{prellberg2004a-a}, we also measured the number of samples
adjusted by the number of their independent growth steps between
$S^{eff} \simeq 4.6 \cdot 10^7$ and $4.4 \cdot 10^8$ ``effective
samples'' at the maximum length. Additionally, for the sake of comparison, we simulated the model
at $p = 0$ by putting the trails on the L-lattice, with $S = 10^6$
iterations we collected $2.2 \cdot 10^9$ samples at the maximum
length, corresponding to $7.4 \cdot 10^7$ effective samples.

\subsection{Specific heat}
\label{sec:specific-heat}

\begin{figure}[ht!]
  \centering
  \includegraphics[width=0.5\columnwidth]{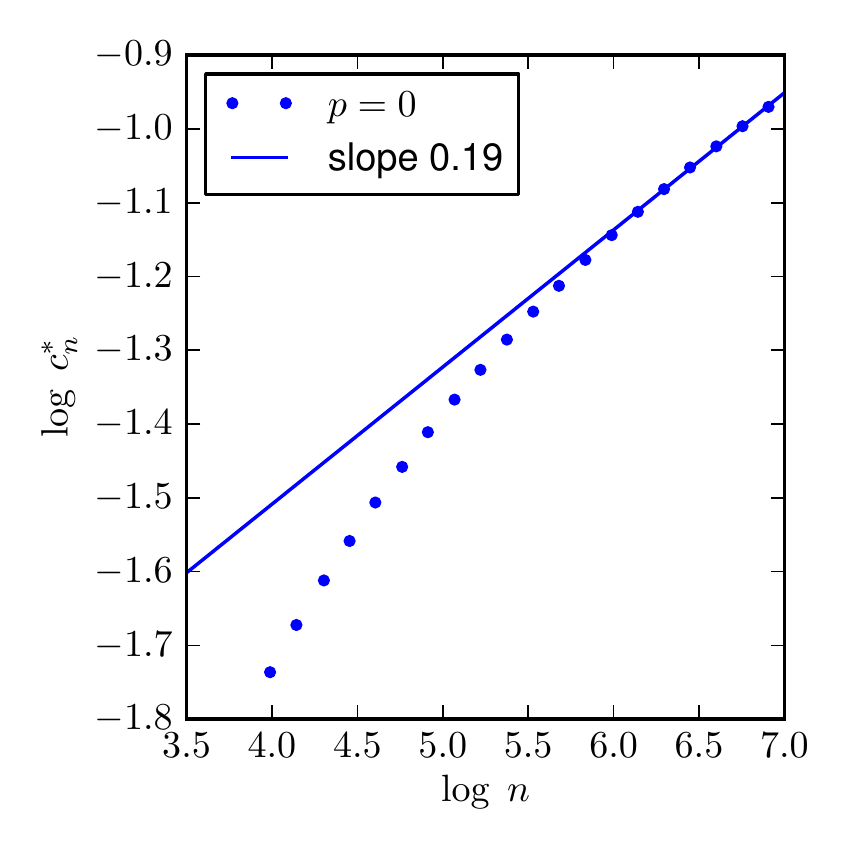}%
  \includegraphics[width=0.5\columnwidth]{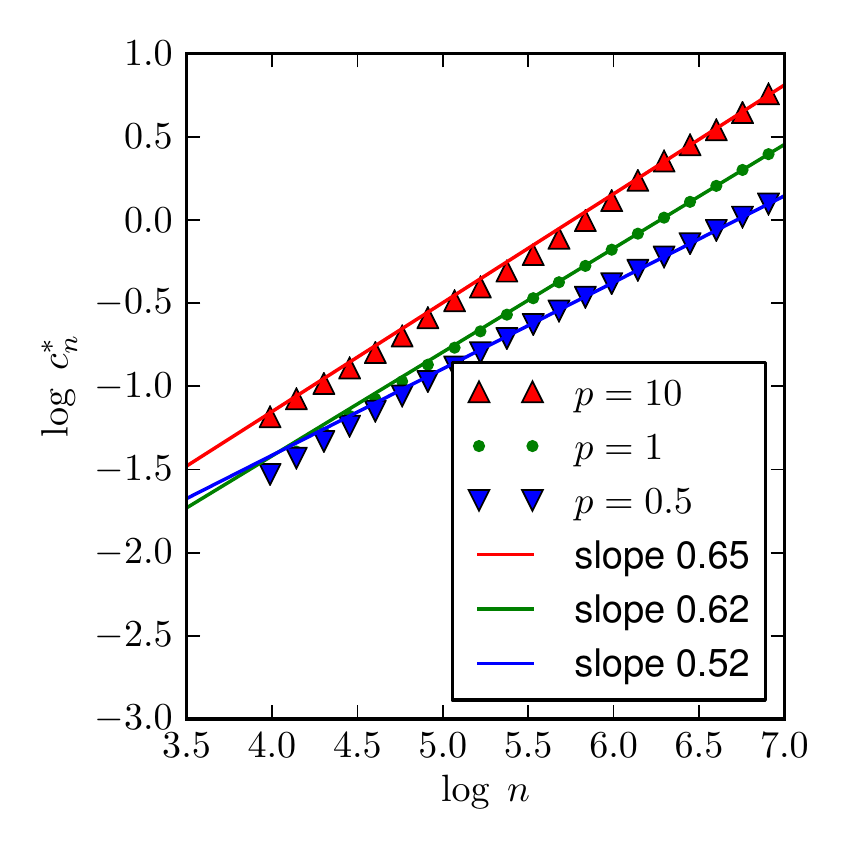}
  \caption{Plot of the logarithm of the peak value of the specific heat versus the logarithm of the length of the
    trails for $p = 0$ (left) and $p = 0.5, 5, 10$ (right). On the right, straight lines fit the data well, whereas
    there is clear curvature in the data on the left.}
  \label{fig:specific-heat}
\end{figure}

We have begun by analysing the scaling of the specific heat by
calculating the location of its peak $\tau_n^* = \argmax_{\tau}\
c_n(\tau)$ and thereby evaluating $c_n^* = c_n(\tau^*_n)$.  In
Figure~\ref{fig:specific-heat}, we plot the peak values of the
specific heat for some of the models we have simulated. The exponent
associated with the peak of the specific heat, see
equation~(\ref{spec-heat-peak-scaling}), is $\alpha\phi$ if the
transition is second order.  For $p=1$ 
we estimate $\alpha\phi\approx 0.64$ from trail lengths up to $1024$,
which is a little less that our previous estimate of $0.68(5)$  based upon
much longer length trails \cite{owczarek2007c-:a}. For $p > 1$ we estimate 
$\alpha\phi \approx 0.74$ at $p = 5$ and $\alpha\phi \approx 0.67$ at $p = 10$,
both of which are compatible with that previous estimate for the ISAT
transition of $0.68(5)$. For $p=0$ we expect that the specific heat does not diverge, 
but at the lengths we consider it is not surprising to measure a weakly increasing specific heat peak with significant curvature in a log-log plot. We can then compare this to the situation when $p=0.5$. Although our estimate of $\alpha\phi \approx 0.52$ is somewhat smaller at $p = 0.5$ than at $p=1$, it is still significantly larger than at $p=0$  and unless one is willing to postulate continuously changing exponents this suggests that 
the entire line up to (but excluding) $p = 0$ is in the ISAT
universality class. There could, of course, be curvature in the plot that we cannot observe at this length scale. However, we reinforce our conclusion by examining the low temperature phase below.

Before we consider the low temperature behaviour of our model, we also observe that the location of the transition in the thermodynamic limit, $\lim_{n\rightarrow\infty} \tau^*_n(p) = \tau_c(p)$, seems to obey
\begin{equation}
1<\tau_c(p)\lesssim 3 \qquad \mbox{ for all } p,
\end{equation}
attaining the value of $\tau_c(p)=3$ when $p=1$. As $p$ increases from zero, $\tau_c(p)$ increases until $p$ reaches some value near one, and then 
decreases again.

\begin{figure}[ht!]
  \centering
  \includegraphics[width=0.5\columnwidth]{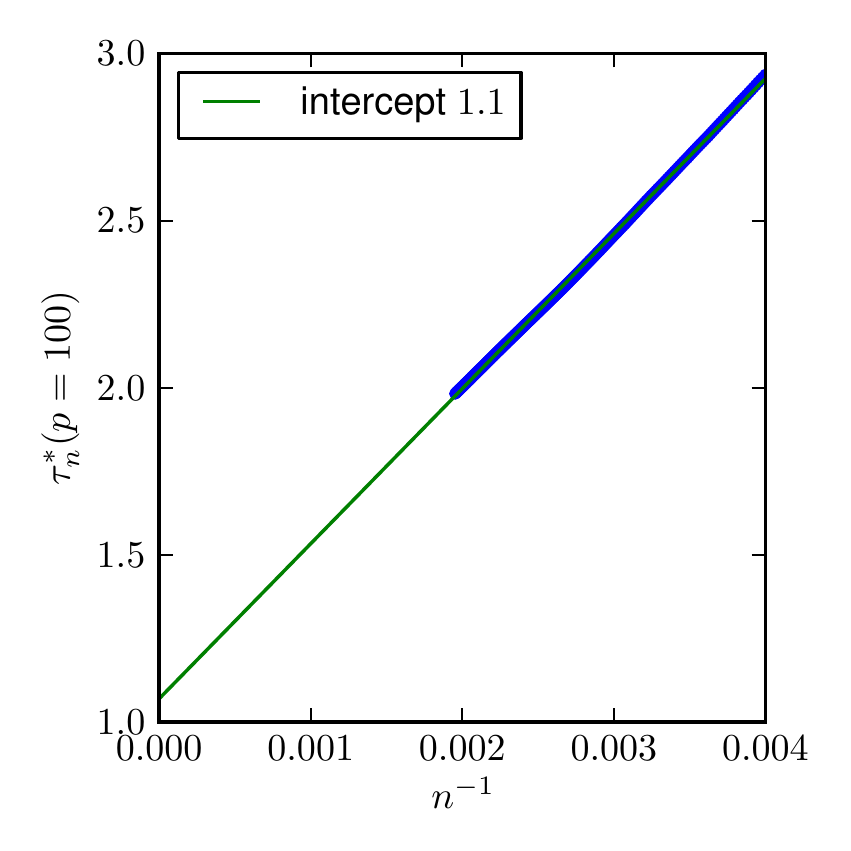}%
  \caption{Plot of the location of the peak of the specific heat for
    $p=100$ versus inverse length $1/n$. A simple linear extrapolation
    gives us the estimate $\tau_c(100) \approx 1.1$.}
  \label{fig:tau_extrap}
\end{figure}

As indicated in Figure~\ref{fig:tau_extrap}, for fixed large values of
$p$, the location of the transition $\tau_c(p)$ is close to, but
larger than one.  We therefore conjecture that $\tau_c(p) \rightarrow
1$ as $p\rightarrow\infty$.  The limit can be understood by the
following ground state-type argument.  For very large $p$ the ground
state depends on whether $\tau$ is smaller or larger than one.  For
$\tau<1$ the ground state is a single straight rod, while for $\tau>1$
it is a configuration that fills every edge of the lattice in order to
maximise doubly visited sites (that is, a set of crossing long rods
that join up on the surface of the polymer).  We point out that such
an argument also naturally leads to a first-order transition at
infinite $p$.

\subsection{Energy distribution}

\begin{figure}[ht!]
  \centering
  \subfigure[$p=5$]{\includegraphics[width=0.5\columnwidth]{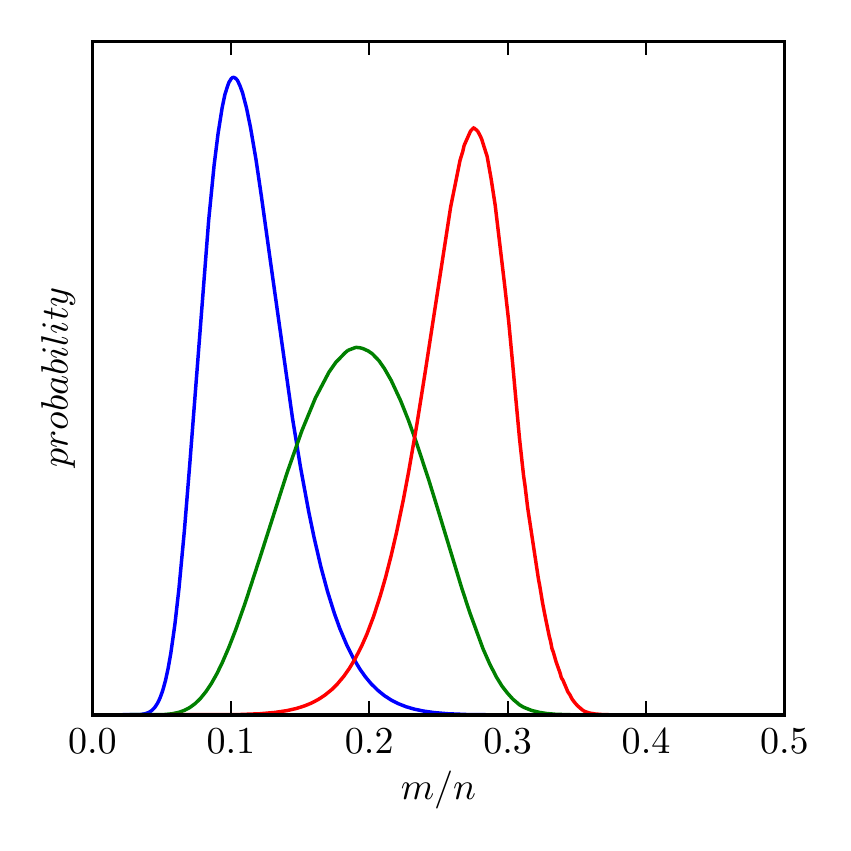}}%
  \subfigure[$p=10$]{\includegraphics[width=0.5\columnwidth]{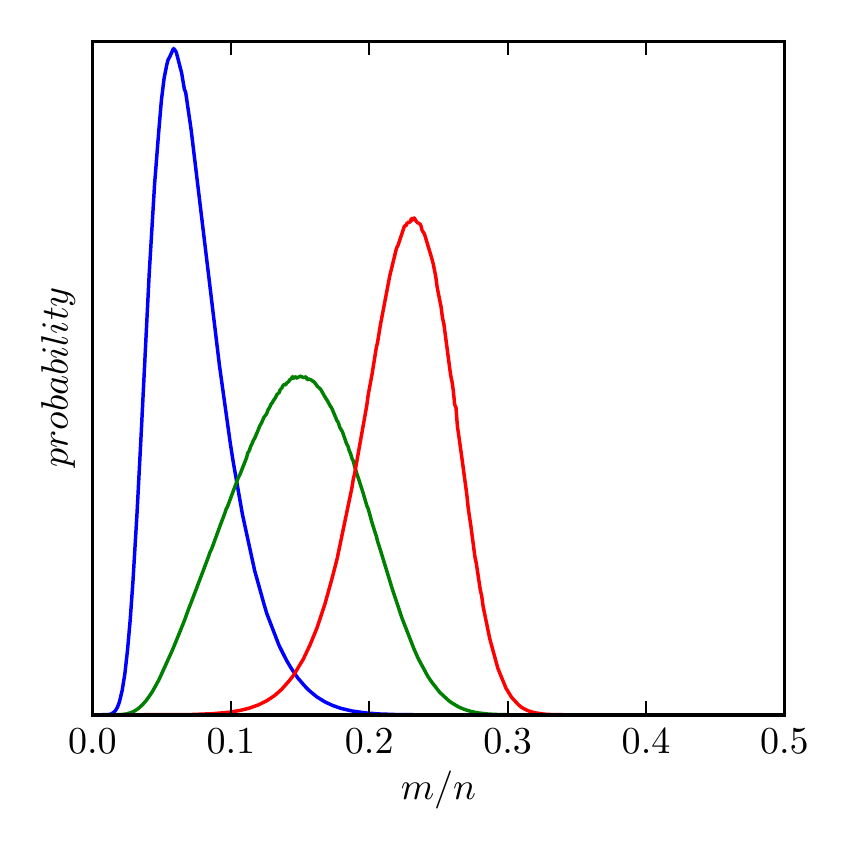}}
  \caption{Energy distribution across the critical region at length $n
    = 1024$ for $p=5$ (a) and $p=10$ (b). For both figures the chosen
    temperatures are $\tau_{\text{left}} = 1.85$,
    $\tau_{\text{middle}} = 1.93$, and $\tau_{\text{right}} =
    2.02$. The weight $\tau_{\text{middle}}$ corresponds to the
    location of the specific heat peak at this length ($\tau^*_n$ in
    the text), while $\tau_{\text{left}}$ and $\tau_{\text{right}}$
    are given by $\tau_{\text{middle}} \pm \frac12 \Delta \tau_n$
    where $\Delta\tau_n$ is the width of the specific heat peak at
    that length as measured at its half height. The distributions are
    normalised to have equal area.}
  \label{fig:energy-distribution}
\end{figure}

We then investigated the claim \cite{foster2011a-a} that for large
finite values of $p$ the collapse transition becomes first-order. We
did so by looking for the presence of bimodality in the energy
distribution. In Figure~\ref{fig:energy-distribution} we plot the
energy distribution in a neighbourhood of the critical point for $p =
5$ and $p = 10$.  We don't find any evidence of a bimodality in either
case.

\begin{figure}[ht!]
  \centering
  \includegraphics[width=0.5\columnwidth]{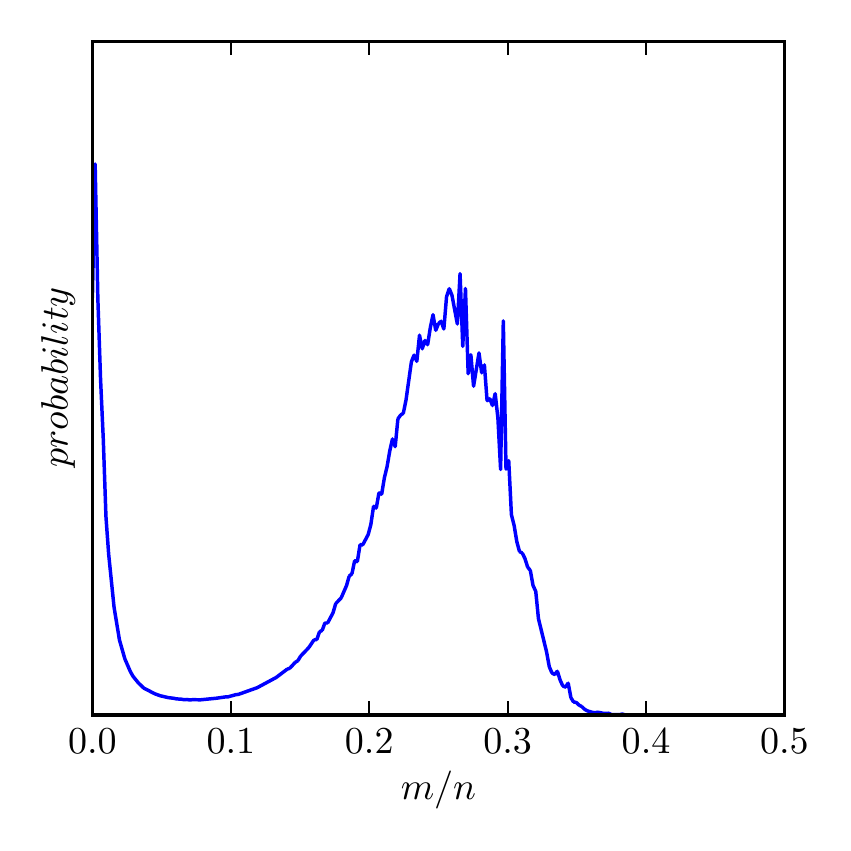}
  \caption{Energy distribution at length $n = 512$ for $p=100$ at
    Boltzmann weight $\tau = 2.01$, chosen so that the heights of the
    two peaks are roughly equal, very close to the specific heat peak
    located at $\tau^* =1.98$.}
  \label{fig:energy-distribution-p100}
\end{figure}

On the other hand, we do
find evidence of a double peak forming at very large values of $p$. In Figure~\ref{fig:energy-distribution-p100} we plot
the energy distribution near the transition for $p=100$, which has a clear double peak.
Given the lengths of trails considered it is not clear immediately whether this is a manifestation of a crossover to an 
infinite $p$ behaviour, which we argued above may well be first-order, or a real change in the order of the collapse 
transition at finite $p$.

\begin{figure}[ht!]
  \centering
  \includegraphics[width=0.5\columnwidth]{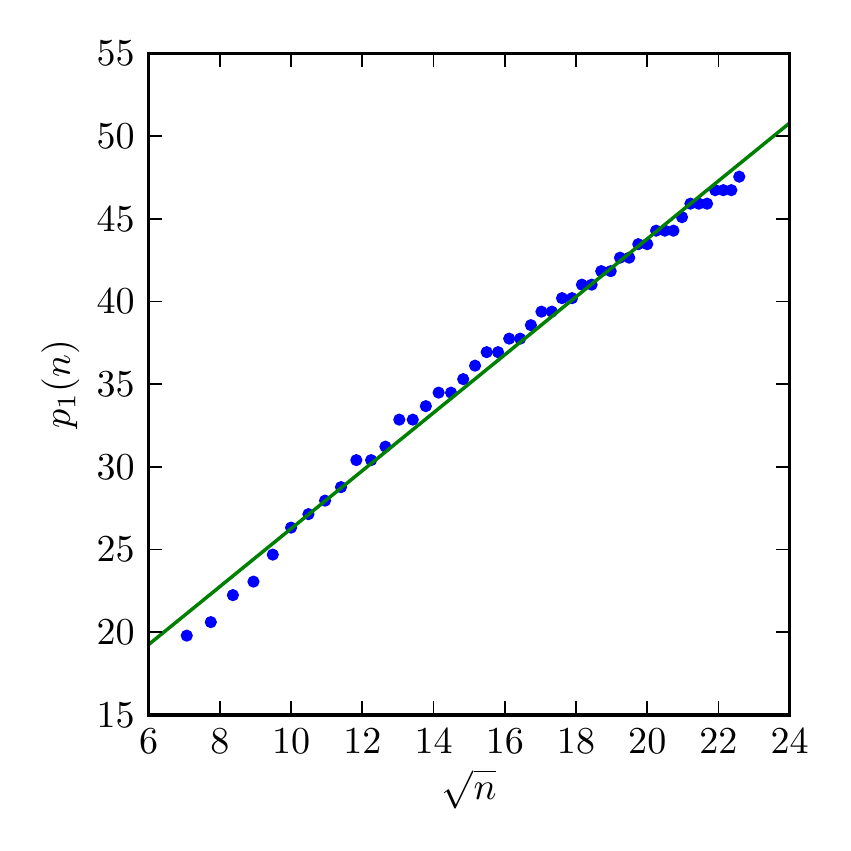}
  \caption{Plot of smallest $p$ value $p_1(n)$ for which an observable
    latent heat is encountered for a finite length trail, versus $\sqrt{n}$.
    This scale is chosen to reflect the possibility of finite-size corrections 
    due to the presence of a surface in the bulk.}
  \label{fig:lat-heat}
\end{figure}

To investigate further we considered the smallest value of $p$ at
which there is a non-zero latent heat at the transition point
$\tau^*_n(p)$ as a function of $n$: we denote this as $p_1(n)$. In
Figure~\ref{fig:lat-heat} we plot $p_1(n)$.

We immediately see that $p_1(n)$ increases in $n$. Hence, if the
transition becomes first-order at some finite value of $p$ for
infinite length trails, this would only occur for very large values of
$p$. In fact, one can see that the increase in $p_1(n)$ is compatible
with a power law such as $n^{1/2}$, which one would expect if the finite-size
corrections were related to surface contributions. Additionally, we have considered
the finite-size latent heat at fixed large $p$. An extrapolation against $n$ is
compatible with a thermodynamic value of zero. This scenario would imply that we
are simply seeing the crossover to the infinite-$p$ behaviour in the
simulations, and that the transition in the thermodynamic limit of
infinite length stays second-order for all finite values of $p$.

\subsection{Low temperature region}
\label{sec:low-temperature}

\begin{figure}[ht!]
  \begin{center}
    \includegraphics[width=0.5\linewidth]{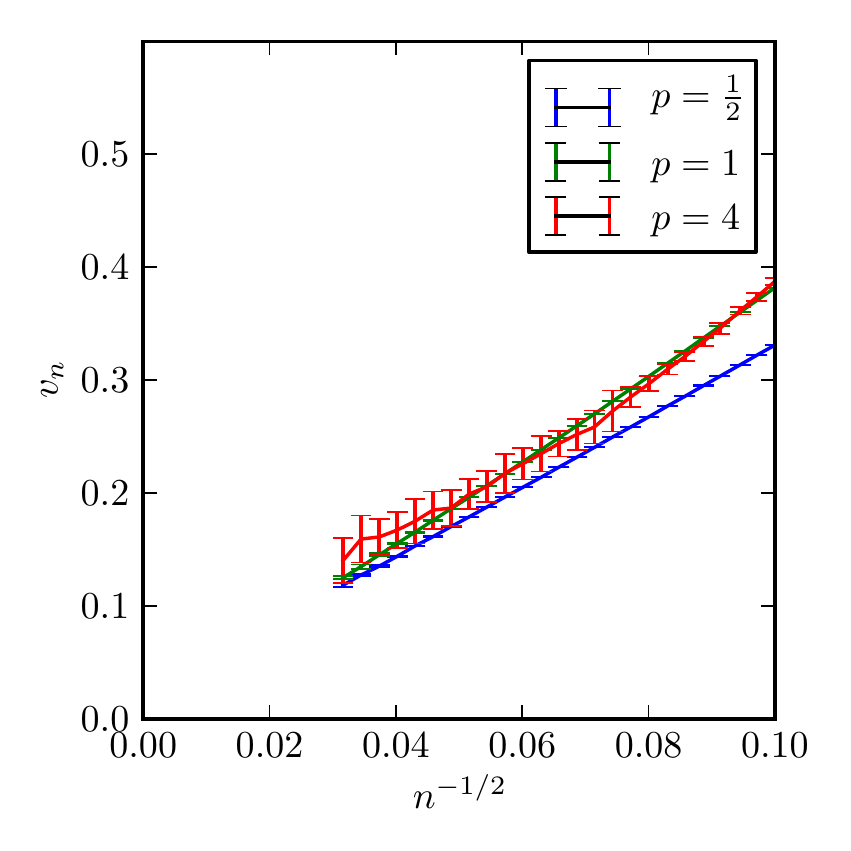}
  \end{center}
  \caption{Plots of $v_n$, the proportion of steps visiting the same
    site once at $\tau = 5$, which is in the low temperature phase,
    against $n^{-1/2}$. The scale $n^{-1/2}$ chosen is the natural low
    temperature scale. }
      \label{fig:low-temperature}
\end{figure}

There is strong evidence that the low-temperature phase of the ISAW
model is a globular phase that is not fully dense, while for
interacting trails the low-temperature phase is maximally dense \cite{bedini2012=a-:a}, i.e.,
the trail fills the lattice asymptotically. Therefore, for trails the
portion of steps not involved with doubly-visited sites should tend to
zero as $n \to \infty$ in this phase.  Following the analysis in
\cite{doukas2010a-:a,bedini2012=a-:a} we measured the proportion $v_n$ of steps not
visiting the same site twice at $\tau=5$, corresponding to a
sufficiently low temperature to be in the collapsed phase.
Figure~\ref{fig:low-temperature} shows a plot of $v_n$ against
$n^{-1/2}$, clearly indicating that $v_n\to0$ within error bars.  This
reinforces our earlier conclusion that for $0< p\leq 1$ the nature of
collapse is that of the standard fully-flexible ISAT model, rather
than changing to the ISAW-like $\theta$-transition that occurs when
$p=0$.

\section{Phase diagram and Conclusions}
\label{sec:pd}

We have studied a generalised model of semi-flexible interacting
trails (SFISAT) by including a stiffness parameter. From our analysis,
we conjecture that the ISAT universality class is unaffected by the
presence of stiffness.  The universality class only changes in the
singular limits of $p\to0$ and $p\to\infty$. In the former limit the
transition is $\theta$-like, whereas the transition turns first-order
in the latter.

While at the lengths considered there are clear bimodal distributions for large values of $p$,
our numerical evidence strongly suggests that these are likely to be finite-size effects associated with
a crossover to a first-order phase transition at infinite stiffness.

\begin{figure}[ht!]
  \centering 
  \begin{tikzpicture}[y=0.75cm]
    \fill[green!10] (0,0) -- (2,0) -- (1.47,5) -- (1.35,6) -- (0,6);

    \fill[fill=blue!10] (2,0) .. controls (2.8,0.5) .. (3,1)
    .. controls (2.65,2) .. (2.13,3) .. controls (1.77,4) .. (1.47,5)
    -- (1.35,6)
    -| (5,0);

    \filldraw[fill=green!10] (2,0) .. controls (2.8,0.5) .. (3,1)
    .. controls (2.65,2) .. (2.13,3) .. controls (1.77,4) .. (1.47,5) -- (1.35,6);

    \draw[black,solid] (1,-0.2) -- (1,0) node[below=2pt] {$1$}
    (2,-0.2) -- (2,0) node[below=2pt] {$2$};
    \draw[black,solid] (-0.2,1) -- (0,1) node[left=2pt] {$1$};

    \fill[blue] (2,0) circle (2pt); \fill[red] (2.96,1) +(-1.5pt,-1.5pt)
    rectangle ++(3pt,3pt); \draw[->] (-0.2,0) -- (5.5,0) node[below]
    {$\tau$}; \draw[->] (0,-0.2) -- (0,6.5) node[left] {$p$};
  \end{tikzpicture}
  \caption{Schematic diagram of the SFISAT parameter space. The filled circle (blue online)
    on the horizontal axes depicts the location of the $\theta$-like transition at $(\tau,p)=(2,0)$, 
    and the filled square (red online) depicts the location of the fully-flexible ISAT transition at $(\tau,p)=(3,1)$.
    The solid line corresponds to an ISAT-like transition. The whole region to the right of the curve, excluding
    the $p = 0$ axis, is maximally-dense.}
  \label{fig:phase-space}
\end{figure}
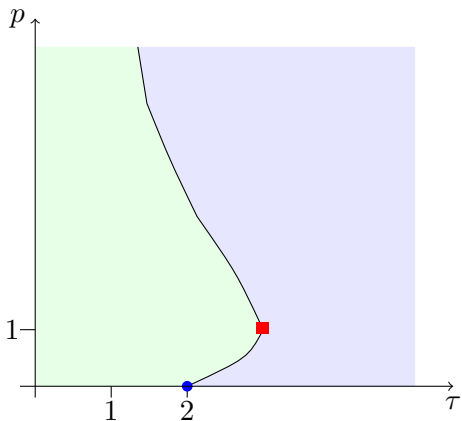

Our results indicate that for all finite values of $p$ the
low-temperature phase is maximally dense as in the fully-flexible ISAT
model.  Hence, unlike the ISAW model, there continues to be only a
single low temperature phase when stiffness is added to ISAT. Putting
the conclusions together suggests the phase diagram shown in
Figure~\ref{fig:phase-space}.

It would be of some interest to examine the effect of stiffness on the canonical ISAT model on the triangular lattice where the low 
temperature phase is globular rather than fully dense as it is on the square lattice considered in this paper.

\section*{Acknowledgements}

Financial support from the Australian Research Council via its support
for the Centre of Excellence for Mathematics and Statistics of Complex
Systems is gratefully acknowledged by the authors. The simulations were performed on the
computational resources of the Victorian Partnership for Advanced
Computing and the University of Melbourne High Performance Computing service. 
A L Owczarek thanks the School of Mathematical Sciences, Queen Mary, University of London for
hospitality.

\end{document}